# The Inherent Geometry of the Nuclear Hamiltonian

Norman D. Cook
Department of Informatics, Kansai University, Osaka 569-1095 Japan

## Abstract

The symmetries inherent to the nuclear version of the Schrödinger wave-equation are identical to the symmetries of a face-centered cubic lattice, as already noted by Wigner in 1937 in the initial development of the independent-particle model (IPM) of nuclear structure. The significance of the identity is that it implies a high-density version of the IPM that has the gross properties of a liquid-drop, rather than a diffuse, chaotic gas of nucleons. As a consequence, all of the predictive strengths of the liquid-drop model (binding energies, radii, nuclear densities, vibrational states, etc.) and the "independent-particle" predictions of the IPM (nuclear spins, parities, magnetic moments, etc.) are retained in the lattice.

## 1 Introduction

Recent indications that random matrix theory [1] – a computational technique heavily used to explain the structure of the atomic nucleus since the 1960s – is invalid [2] have motivated a re-evaluation of the standing-wave lattice model of nuclear structure. The idea that the nucleus is *not* a chaotic "Fermi gas", but rather a lattice of nucleons was first described by Eugene Wigner [3], who noted that: "The quantum numbers are all half-integers [whose] combinations form a face centered lattice…" (p. 106). As shown in Figure 1 of Wigner's 1937 paper in ***Physical Review*** (reproduced here as Figure 1A), the quantum number symmetries of the nuclear Hamiltonian can be expressed as the shells and layers inherent to a close-packed lattice. The geometry of the sequential shells and subshells of the harmonic oscillator is more easily seen in 3D structures and in dynamic computer graphics (Figure 1B, C), but the original insight of the lattice representation of nuclear quantum space was Wigner's.

    The terminology and full set of quantum numbers for the nuclear Hamiltonian were subsequently revised in light of spin-orbit coupling [4], but the discovery of the IPM in the 1930s was explicitly cited in awarding Wigner one-half of the 1963 Nobel Prize in Physics, while the remaining half went to the inventors of the shell model. The idea that the nucleus is a face-centered-cubic (fcc) lattice of nucleons was later developed explicitly as a model of nuclear structure by others, and has been described as a "dynamic lattice" [5], an "array of Gaussian probability clouds" [6], a "standing wave of nucleons" [7~9], a "condensate of correlated quarks" [10] or a "lattice gas" [11~13].

    Despite Wigner's specification of the lattice geometry, the quantal symmetries of the nucleus have generally been interpreted in terms of an abstract momentum-space rather than coordinate-space. The emphasis on momentum-space was of course consistent with shell model assumptions concerning a central, nuclear potential-well and the chaotic movement of "point" nucleons within the potential-well, but both of these assumptions have remained problematical in nuclear structure theory [14]. Specifically, (i) the presumption of a long-range "effective" nuclear potential-well stands in contradiction to the short-range, realistic, nuclear force known

from nucleon-nucleon scattering experiments [15]; (ii) unlike electrons, nucleons have measurable electrostatic and/or magnetic RMS radii of ~0.9 fermi [16] that make intranuclear "orbiting" unlikely and the nucleon's mean-free-path short (~3 fm), e.g., [17]; and (iii) the presumed chaotic motion of bound nucleons has been found to be incorrect in the only direct test of random matrix theory [2].

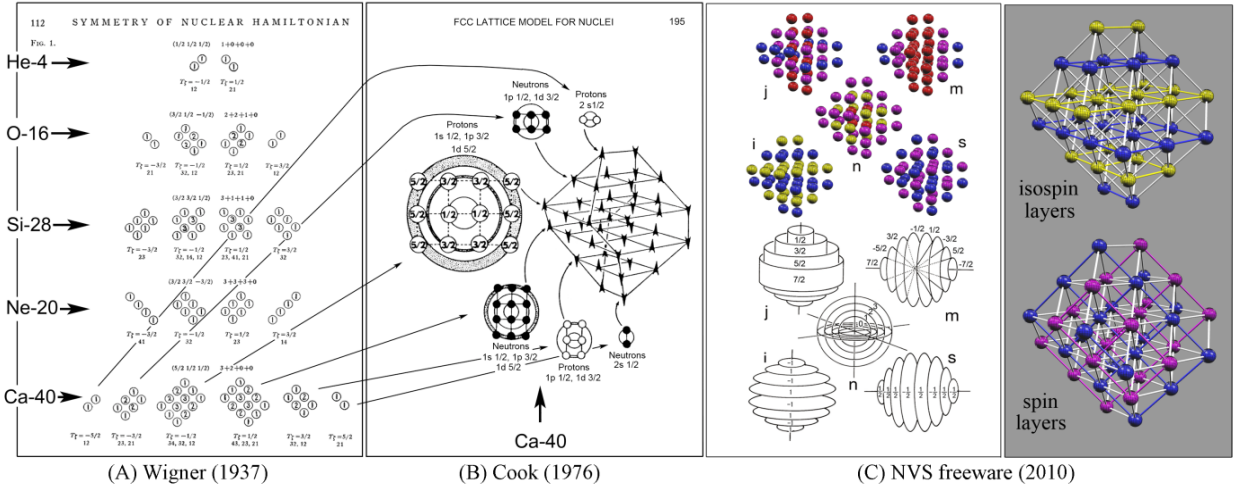

Figure 1: Illustration of the lattice symmetries of the nuclear Hamiltonian: (A) as shown in Figure 1 from Wigner [3]; (B) the same lattice substructure in the fcc nuclear model as shown in Figure 3 from Cook [7]. Note the precise identity between the unstacked layers of Wigner's depiction of $^{40}$Ca (bottom row in A) and the layers of the reconstructed 3D model of $^{40}$Ca (B). (C) shows the same geometry with separate depiction of the nucleon quantum numbers for the 40 nucleons of $^{40}$Ca and close-ups of spin- and isospin-layering, as displayed in the Nuclear Visualization Software [14,20].

Specifically, in a careful analysis of high-quality cross-section measurements on two isotopes of Platinum, Koehler and colleagues [2] have demonstrated that the assumption of a random matrix in calculating neutron width resonances is *invalid*. Since there are no grounds to expect that the quantum mechanics of Platinum nuclei would differ substantially from the quantum mechanics of other nuclei, they have commented that "violation of this assumption could have far-reaching consequences" for nuclear theory, in general, and, moreover, that "there is no viable model of nuclear structure that could explain these [results]" [18]. To the contrary, I maintain that a standing-wave lattice model of the nucleus precisely reproduces the regularities of the IPM that have been confirmed by many decades of low-energy nuclear experimentation, without relying on the chaotic motion of nucleons, as postulated in random matrix theory.

## 2 The Identity between Nuclear States and Lattice Geometry

The successes of the conventional IPM are based on the nuclear Hamiltonian, where all possible nucleon states are given by the nuclear version of the Schrödinger equation:

$$\Psi_{n,\, j(l+s),\, m,\, i} = R_{n,\, j(l+s),\, i}(r)\, Y_{m,\, j(l+s),\, i}(\theta, \phi) \qquad (1)$$

Of interest with regard to the inherent geometry of the nucleus are the relationships among the quantum numbers and the patterns of occupancy of nucleon energy states that are implied.

The universally-acknowledged strength of the IPM (ca. 1950) lay in the fact that the state of each "independent" nucleon in the model is specified by a unique set of quantum numbers (*n, j, m, l, s, i*), allowing for an explanation of *nuclear* states as the summation of *nucleon* states. The range of values that the quantum numbers can take is known both theoretically and experimentally to be:

*n* = 0, 1, 2, …     *j* = 1/2, 3/2, 5/2, …, (2*n*+1)/2     *m* = -*j*, …, -5/2, -3/2, -1/2, 1/2, 3/2, 5/2, …, *j*

*s* = 1/2, -1/2 (up, down)     *i* = 1, -1 (protons, neutrons)     *parity* = even, odd and this pattern gives each nucleon a unique set of quantum numbers. Together with the Schrödinger equation itself, these equations are essentially a concise statement of the quantum mechanics of the nucleus. Quantum numbers *j*, *s*, *i* and *parity* are observable quantities, and have allowed for innumerable predictions concerning the known 1800+ isotopes and their many excited states. While the liquid-drop model (LDM), the cluster models and more than 30 (!) other nuclear models [19] are still used to explain the diversity of nuclear phenomena, the quantum mechanical successes of the IPM have established it as the central paradigm of nuclear structure theory since the 1950s.

The physical interpretation of the fcc lattice is an interesting problem and remains debatable (see below), but the precise identity between the nuclear Hamiltonian and the fcc lattice is unambiguous [Eqs. (2)~(10)]. As a consequence, the Cartesian coordinates for each nucleon can be used to define its quantum numbers:

$$n = (|x| + |y| + |z| - 3) / 2 \qquad (2)$$
$$j = (|x| + |y| - 1) / 2 \qquad (3)$$
$$m = s * |x| \qquad (4)$$
$$s = (-1)^{(x-1)/2} / 2 \qquad (5)$$
$$i = (-1)^{(z-1)/2} \qquad (6)$$
$$parity = \text{sign}(x*y*z) \qquad (7)$$

Or, vice versa, the unique quantal state of each nucleon can be used to define its Cartesian coordinates in the lattice:

$$x = |2m|(-1)^{(m-1/2)} \qquad (8)$$
$$y = (2j+1-|x|)(-1)^{(i+j+m+1/2)} \qquad (9)$$
$$z = (2n+3-|x|-|y|)(-1)^{(i+n-j)} \qquad (10)$$

Going either from quantum mechanics to the lattice or vice versa, the *known* sequence and relations among quantum numbers and the *known* occupancy of protons and neutrons in the *n*-shells and *j*- and *m*-subshells are identical in both descriptions.

As shown most clearly in Figure 1C, the *n*-, *j*- and *m*-shells have spherical, cylindrical and conical symmetries, respectively, while *s*- and *i*-values produce orthogonal layering. Noteworthy is the fact that a nucleon's principal quantum number *n* is dependent on the nucleon's distance from the center of the system, and its total angular momentum value *j* is dependent on the nucleon's distance from the nuclear spin axis. Moreover, the lattice implies that nuclear parity can be defined simply as the sign of the product of nucleon coordinates, resulting in the known alternations of positive and negative parity with sequential *n*-shells of the isotropic harmonic oscillator (Table 1). The validity of Eqs. (2)~(10) can be verified nucleon-by-nucleon (Figure 2), and the entire pattern – found in both the IPM and in the fcc lattice built from a central tetrahedron – can be summarized as in Table 1.

It is worth noting that Eqs. (2)~(10) do *not* imply a return to classical mechanics, but they do imply a specific, non-random, non-chaotic geometry of the quantized substructure of the nucleus. The geometrical pattern of principal quantum number, n, shells produces literal shells; the subshells of total angular momentum, j, show j-values to be dependent on the nucleon's distance from the nuclear spin-axis. And the spin- and isospin-layering gives the lattice the properties of an antiferromagnet, with a small attractive magnetic force working between nearest neighbors in each isospin layer. From theoretical work on neutron stars, the magnetic attraction in an fcc lattice with isospin-layering has previously been shown to be the reason why this particular close-packed lattice is the lowest-energy solid-phase configuration of nucleons (N=Z), possibly realized in the crust of neutron stars [21].

Table 1: Occupancy of the *n*-shells and *j*- and *m*-subshells in the IPM and lattice models.

| parity $\pi = \text{sign}(x \ast y \ast z)$ | n-shells $n = (\|x\|+\|y\|+\|z\|-3)/2$ | | | | | | j-subshells $j = (\|x\|+\|y\|-1)/2$ | | | | | | | m-subshells $m = \|x\|/2$ | | | | | | Total |
|---|---|---|---|---|---|---|---|---|---|---|---|---|---|---|---|---|---|---|---|---|
| | 0 | 1 | 2 | 3 | 4 | 5 | $\frac{1}{2}$ | $\frac{3}{2}$ | $\frac{5}{2}$ | $\frac{7}{2}$ | $\frac{9}{2}$ | $\frac{11}{2}$ | $\frac{13}{2}$ | $\frac{1}{2}$ | $\frac{3}{2}$ | $\frac{5}{2}$ | $\frac{7}{2}$ | $\frac{9}{2}$ | $\frac{11}{2}$ | $\frac{13}{2}$ | |
| + | 2 | | | | | | 2 | | | | | | | 2 | | | | | | | 2** |
| | | | | | | | | 4 | | | | | | 2 | 2 | | | | | | 6* |
| − | | 6 | | | | | 2 | | | | | | | 2 | | | | | | | 8** |
| | | | | | | | | | 6 | | | | | 2 | 2 | 2 | | | | | 14* |
| | | | | | | | | 4 | | | | | | 2 | 2 | | | | | | 18 |
| + | | | 12 | | | | 2 | | | | | | | 2 | | | | | | | 20** |
| | | | | | | | | | | 8 | | | | 2 | 2 | 2 | 2 | | | | 28** |
| | | | | | | | | | 6 | | | | | 2 | 2 | 2 | | | | | 34* |
| | | | | | | | | 4 | | | | | | 2 | 2 | | | | | | 38 |
| − | | | | 20 | | | 2 | | | | | | | 2 | | | | | | | 40* |
| | | | | | | | | | | | 10 | | | 2 | 2 | 2 | 2 | 2 | | | 50** |
| | | | | | | | | | | 8 | | | | 2 | 2 | 2 | 2 | | | | 58* |
| | | | | | | | | | 6 | | | | | 2 | 2 | 2 | | | | | 64* |
| + | | | | | | | | 4 | | | | | | 2 | 2 | | | | | | 68* |
| | | | | | 30 | | 2 | | | | | | | 2 | | | | | | | 70* |
| | | | | | | | | | | | | 12 | | 2 | 2 | 2 | 2 | 2 | 2 | | 82** |
| | | | | | | | | | | | 10 | | | 2 | 2 | 2 | 2 | 2 | | | 92* |
| | | | | | | | | | | 8 | | | | 2 | 2 | 2 | 2 | | | | 100 |
| | | | | | | | | | 6 | | | | | 2 | 2 | 2 | | | | | 106 |
| | | | | | | | | 4 | | | | | | 2 | 2 | | | | | | 110 |
| − | | | | | | 42 | 2 | | | | | | | 2 | | | | | | | 112 |
| | | | | | | | | | | | | | 14 | 2 | 2 | 2 | 2 | 2 | 2 | 2 | 126** |
| | | | | | | | | | | | | 12 | | 2 | 2 | 2 | 2 | 2 | 2 | | 138 |

Double asterisks (**) denote harmonic oscillator shell/subshell closures suggestive of "magic" stability; single asterisks (*) denote shell/subshell closures for which there is weaker evidence for "magic" stability [14]. The sequence of j-subshells within the n-shells is variable ("configuration mixing"), nucleus by nucleus – leading to some variation concerning the closure of shells and subshells in the shell model.

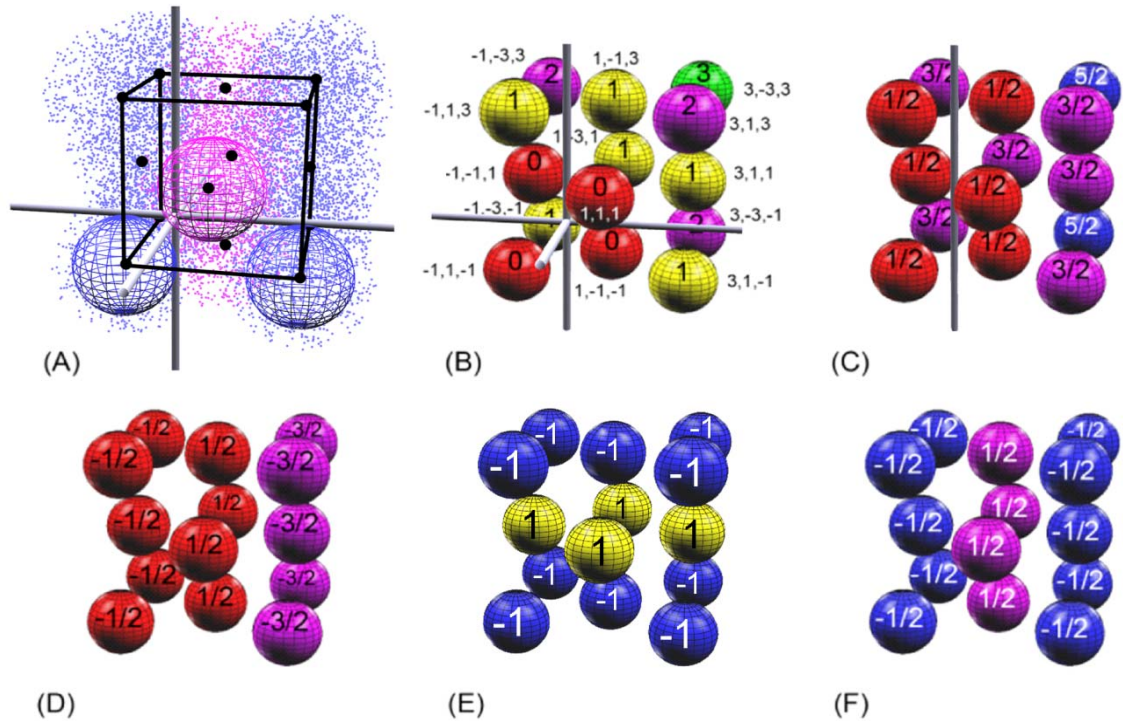

Figure 2: Six depictions of the 14-nucleon "unit structure" of the fcc lattice. The unit corresponds to a highly unstable isotope of Beryllium, $^{14}$Be, and is shown here only to illustrate the precise geometry of quantum numbers in the lattice. (A) shows the known nuclear density (0.17 nucleons/fm$^3$) and Gaussian "probability clouds" of the 14 "point" nucleons, with the 90% probability wire-spheres illustrating the known dimensions of the nucleon itself (r=0.9 fm). (B)~(F) illustrate the assignment of quantum numbers depending solely on nucleon lattice coordinates. (B) Principal quantum number *"n"* is dependent on the nucleon's position relative to all three axes (red, *n*=0; yellow, *n*=1; purple, *n*=2; green, *n*=3). (C) Total angular momentum number *"j"* (=|*l+s*|) is dependent on the nucleon's distance from the nuclear spin-axis (red, *j*=1/2; purple, *j*=3/2; blue, *j*=5/2). (D) Azimuthal quantum number *"m"* is dependent on the nucleon's distance from the YZ-plane and the nucleon's spin value (red, *m*=|1/2|; purple, *m*=|3/2|). (E) The isospin quantum number *"i"* alternates in layers along the Z-axis (yellow, *i*=1; blue, *i*=-1). (F) The spin quantum number *"s"* alternates in layers along the X-axis (purple, *s*=1/2; blue, *s*=-1/2). The complex structures of larger nuclei are easily examined using software designed for that purpose [20] (available at: www.res.kutc.kansai-u.ac.jp/~cook/nvs).

The mathematical identity between the IPM and the fcc lattice leads to resolution of well-known paradoxes in nuclear structure theory (concerning nuclear densities, nuclear surface properties, alpha-particle clustering, nuclear force dimensions, asymmetrical nuclear fission, etc. [14]). Despite first impressions, the switch from a "chaotic gas" of nucleons to the symmetries of a quantized lattice leaves the IPM completely intact (where the quantum numbers of nucleons are dependent on the nucleon's position within the lattice), while producing a realistic, high-density nuclear core texture akin to the LDM.

# 3 Implications

The significance of the isomorphism between the IPM and the lattice lies in the fact that, if we know the IPM quantal structure of a nucleus, then we also know its spatial structure, or vice versa. The only structural uncertainty in **both** models is that only the quantum number characteristics of the last-odd proton and/or last-odd neutron are known unambiguously from experiment. Even-Z and even-N nuclei are assumed to have paired valence nucleons, differing only in spin, and the core nucleons are assumed to have the same IPM characteristics as known from smaller (odd-Z and/or odd-N) nuclei. Both of these latter assumptions are generally well-justified, but there are in fact many known cases of "intruder states" and "configuration-mixing" in which the default IPM nucleon build-up sequence is not followed (Table 1).

Among the "far-reaching consequences" [2] of the possible demise of nuclear applications of random matrix theory is the reconciliation of the three main models of modern nuclear structure theory within the lattice. Specifically, a high-density nuclear interior dominated by nearest-neighbor nucleon interactions, as postulated in the LDM, is reproduced in the lattice. By discarding the IPM's convenient fiction of a long-range "effective" nuclear force, LDM predictions concerning nuclear binding energies, densities, radii, surface effects, low-lying vibrational states, etc. are retained within the lattice representation of nucleon states. In this view, the "independent" properties of each individual nucleon are defined by the particle's position within the collective, but the macroscopic properties of the nuclear "collective" are better described by the analogy with a liquid-drop than a Fermi gas.

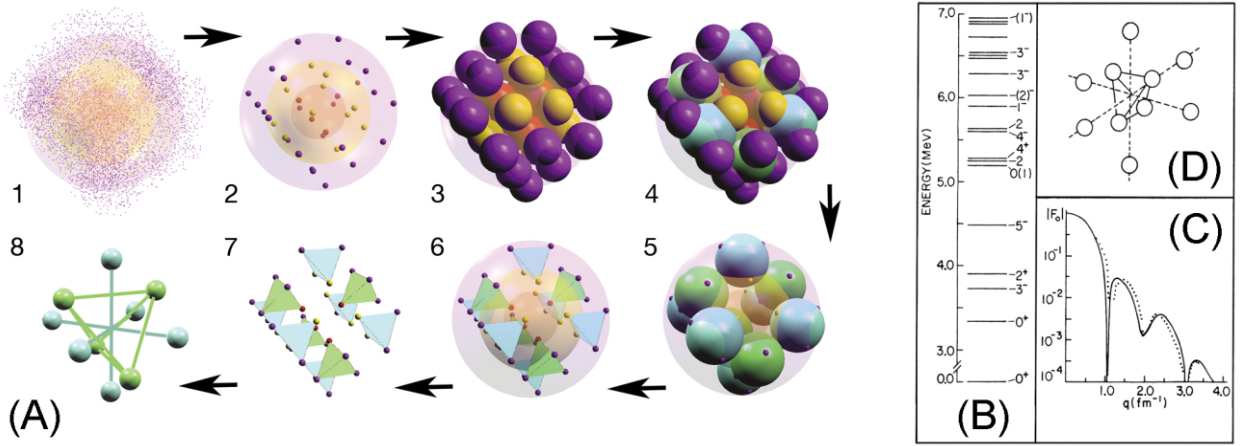

Figure 3: The alpha-particle texture inherent to the fcc lattice. (A) shows 8 depictions of the $^{40}$Ca nucleus in the lattice model: (1) all nucleons depicted as probability clouds, (2) nucleons depicted as point particles in three distinct *n*-shells, (3) nucleons with realistic dimensions (r=0.9 fm), (4) all 40 nucleons grouped into alpha clusters, (5) nucleons reduced in size to emphasize the cluster structure, (6) alphas depicted as tetrahedra (Z=2, N=2) within the *n*-shells, (7) alpha tetrahedra only, (8) the geometry of the alphas. Successful predictions of the alpha-particle model concerning (B) excited states and (C) the electron form-factor for $^{40}$Ca in (c). The same alpha geometry in the cluster model (D) (e.g., ref. [22, 23] is also found in the lattice model.

Conversely, the main *difference* between the conventional IPM and the lattice model lies in their implications concerning the local substructure within the nucleus. The IPM maintains that substructure is a consequence of energy gaps in a ***long-range***, "effective" nuclear potential-well, whereas the lattice model views the same configuration of quantum states as a 3D standing-wave held together by a realistic, ***short-range*** nuclear force, with substructure determined by local nucleon-nucleon interactions, as described in the LDM. The lattice model therefore has properties similar to both the IPM and the LDM, but the lattice has additional substructure not found in either a liquid-drop or a nucleon "gas" of independent particles. The most interesting example is the lattice geometry of $^{40}$Ca, which has an inherent subgrouping of 10 alpha-particle clusters that is identical to the alpha structure postulated in the cluster depiction of this nucleus [22,23] (Figure 3).

Although the cluster model remains a minority concern within nuclear structure theory [24~26], its successes are not easily explained within the framework of either a liquid-drop or a gaseous nuclear interior, but find an unforced interpretation within the lattice model. Specifically, in both the cluster model and the lattice model, the 10 alpha particles of $^{40}$Ca have a geometry that can be described as an octahedron of 6 alphas lying outside of a tetrahedron of 4 alphas – a geometrical configuration that allows for prediction of excited states and the electron form factor for this nucleus (Figure 3 C, D). That alpha structure is found within the LDM-like, uniform density of the lattice and simultaneously exhibits the three doubly-magic closed shells at 4, 16 and 40 nucleons (Figure 3A).

# 4 Conclusions

The identity between the symmetries of the Schrödinger equation (~IPM) and the lattice is unambiguous [Eqs. (2)~(10)], but uncertainty remains concerning its physical interpretation. Inevitably, all of the controversial conceptual issues of quantum theory (e.g., ref. [27]) (debated to a stalemate in the 1920s and 1930s by the giants of quantum physics; for a brief summary of the fifteen "common" interpretations of quantum mechanics, see ref. [28]) are again raised concerning the spatial implications of the Schrödinger equation, the wave-particle duality, and the stochastic nature of fundamental reality. However those issues may eventually be resolved, Wigner's long-overlooked fcc representation of nuclear quantum space provides interesting possibilities for a return to realistic discussions of the coordinate-space structure of the nucleus.